\documentstyle[12pt,amssymb,epsf]{article}
\newcommand{\PRA}[1]{Phys.\ Rev.\ A {\bf #1}}

\newcommand{\PRE}[1]{Phys.\ Rev.\ E {\bf #1}}
\newcommand{\PRL}[1]{Phys.\ Rev.\ Lett.\ {\bf #1}}

\newcommand{\JPA}[1]{J.\ Phys.\ A {\bf #1}}

\newcommand{\JSP}[1]{J.\ Stat.\ Phys.{\bf #1}}

\title{ 
Phase separation in disordered exclusion models}
\author{Joachim Krug\\
{\small Fachbereich Physik, Universit\"at GH Essen, D-45117 Essen, Germany}}
\date{\today}

\begin{document}
\maketitle
\begin{abstract}
The effect of quenched disorder in the one-dimensional asymmetric exclusion
process is reviewed. Both particlewise and sitewise disorder generically
induces phase separation in a range of densities. In the particlewise
case the existence of stationary product measures in the homogeneous
phase implies that the critical density can be computed exactly,
while for sitewise disorder only bounds are available. The coarsening
of phase-separated domains starting from a homogeneous initial condition
is addressed using scaling arguments and extremal statistics considerations.
Some of these results have been obtained
previously in the context of directed polymers subject to columnar disorder.
\end{abstract}

\section{Introduction}

The one-dimensional
asymmetric simple exclusion process (ASEP) was introduced by Spitzer
in 1970 as an example of an interacting stochastic process 
\cite{spitzer,liggett,liggett99}.
In the probabilistic community it has been widely used for rigorous studies of
the emergence of hydrodynamic behavior from stochastic microscopic
dynamics \cite{spohn,kipnis99}. Already thirty years ago similar models
were considered in the context of biopolymerization \cite{schuetz97},
while recent applications have focused on the problem of vehicular
traffic flow \cite{nagel96}. The interest of statistical physicists
has been further fueled by the discovery of boundary-induced phase
transitions \cite{krug91,schuetz93,hakim93} as well as the relations
to interface growth and directed polymers in random media \cite{ks91,hhz95}.
In short, the ASEP is a generic model of driven single file transport which
combines utmost simplicity with a remarkable richness of behaviors. 

Figure 1 illustrates the model. Particles
occupy the sites of a one-dimensional lattice subject to the simple exclusion
rule (at most one particle per site). In an infinitesimal time interval 
$dt$ particle $i$ at site $x_i$ attempts a jump to the right (left)
with probability $p dt$ ($q dt$). The jump succeeds if the neighboring 
site is empty and is suppressed otherwise. In general the jump rates
$p$ and $q$ may depend on both the particle label $i$ and the position
$x$ on the lattice. 
In much of the paper I
will restrict myself to the {\em totally} asymmetric case
$q = 0$. 

In the present article I want to address the effects
that {\em quenched disorder} in the jump rates has on the behavior of the
ASEP. Disorder effects can be quite dramatic in one-dimensional single file
systems, as is evidenced by the everyday experience with platoons
and traffic jams caused by slow vehicles, accidents
or road construction on highways \cite{newell,bennaim94,krug97,csahok94,knospe97}.  
Also in the context of driven transport on biomolecules a certain amount
of disorder seems unavoidable \cite{harms97}. 

It is natural to distinguish between {\em particlewise disorder}
with $p = p_i$, $q = q_i$ independent of $x$, and 
{\em sitewise disorder} with  
$p = p(x)$, $q = q(x)$ independent of $i$. 
Both particlewise and sitewise disorder generically induces 
{\em phase separation} in the sense that, for global particle densities
$\rho$ in a certain interval $[ \rho_c^-, \rho_c^+]$, the system
breaks up into regions of density $\rho_c^-$ and  $\rho_c^+$ separated
by sharp density discontinuities (``shocks''). These shocks are typically
associated with bottlenecks, i.e. slow particles or slow sites in the
particlewise and sitewise cases, respectively. If the system is started
from a homogeneous initial condition, the average size 
$\xi$ of phase separated regions grows as a power law
\begin{equation}
\label{xi}
\xi(t) \sim t^{1/z},
\end{equation}
defining a dynamic exponent $z$; an example of the time evolution in
the particlewise case is shown in Figure 2. Two kinds of questions
will therefore be asked in the following: First, how can the density
interval $[ \rho_c^-, \rho_c^+]$ of phase separation be determined?
Second, what is the value of the dynamic exponent, and how does it
depend on the distribution of the disordered jump rates?

For particlewise disorder a number of exact analytic results are
available \cite{benjamini96,kf,evans96,evans97,timo99} which
have been reviewed elsewhere \cite{krug97}. This case will
therefore only be briefly summarized in Section \ref{Particle}.
The more difficult problem of sitewise disorder has been studied
numerically by Tripathy and Barma \cite{tripathy98} and
others \cite{csahok94,bengrine99}, but little is known analytically.
In Section \ref{Site} some progress in this direction will
be reported. Specifically, I derive a rigorous bound on the critical
densities based on the results for the particlewise case, and
obtain predictions for the coarsening behavior for various
types of disorder distributions. The relation to directed polymers
in random media is briefly discussed in Section \ref{DP}, and 
some conclusions and open questions are formulated in Section 
\ref{Conclusions}.

\section{Particlewise disorder}
\label{Particle}

\subsection{Steady state and critical density}

For particlewise disorder the configurations of the system are most
naturally described in terms of the {\em headways} $u_i = 
x_{i+1} - x_i - 1$ in front of the particles. The key simplifying
feature is that different headways become statistically independent
in the steady state, with a geometric distribution \cite{benjamini96,evans96}
\begin{equation}
\label{Pi}
P_i(u) = (1 - \alpha_i) \alpha_i^{u}
\end{equation}
for the headway in front of particle $i$. In the totally asymmetric case
the parameters $\alpha_i$ are determined by the jump rates $p_i$
through the simple relation \cite{kf,evans96}
\begin{equation}
\label{alphai}
\alpha_i = v/p_i
\end{equation}
where $v$ is the (common) mean speed of the particles in the steady 
state. Eq.(\ref{alphai}) expresses the plausible fact that the 
headways in front of slow particles are larger than in front of 
fast ones. The geometric distribution (\ref{Pi}) remains valid in
the case of partial asymmetry, but then (\ref{alphai}) is replaced
by a more complicated relation 
\cite{benjamini96,evans96}. The steady state distribution for
the totally asymmetric model with parallel update has a similar form
\cite{evans97}.

In the following we consider the totally asymmetric case and
take the $p_i$ to be independent random
variables with a probability density $f(p)$ supported on the interval
$[c,1]$, with a minimal speed $c$ bounded away from zero. Since particles
cannot pass each other, it is clear that the steady state speed $v$ 
in an infinite system cannot exceed $c$. To compute it, one determines
the mean headway in front of particle $i$ from (\ref{Pi}) and performs
the disorder average. In a system of density $\rho$ the resulting 
average headway must be $(1 - \rho)/\rho$. This yields the implicit 
equation
\begin{equation}
\label{v}
\rho  = \left[ 1+ v \int_c^1 \frac{dp \; f(p)}{p - v} \; \right]^{-1}
\end{equation}
for the speed as a function of density. Two cases are to be distinguished.
If the integral on the right hand side of (\ref{v}) diverges in the limit
$v \to c$, then $v(\rho) < c$ for all $\rho > 0$. In this case the
$\alpha_i$ in (\ref{Pi}) are bounded away from unity for all $i$, the 
headway distributions are normalizable, and the system
remains homogeneous. If, on the other hand, the integral remains finite
in this limit, then the right hand side of (\ref{v}) evaluated
at $v = c$ defines a critical density $\rho_c$ such that 
$v(\rho) \equiv c$ in the entire interval $[0, \rho_c]$. 
For the slowest particles with $p_i \approx c$ this implies that
the headway distributions (\ref{Pi}) are no longer normalizable.
Large gaps appear in front of these particles, and the faster particles
form platoons behind them, a phenomenon familiar from vehicular traffic
on country roads \cite{krug97,newell,bennaim94}. 
The system {\em phase separates} into
regions of density $\rho_c^- = 0$ (the gaps) and regions of  density
$\rho_c^+ = \rho_c$ (the platoons). 

It is evident from (\ref{v}) that the condition for phase separation
translates into a condition on the behavior of 
the disorder distribution $f(p)$ near $p = c$.
Introducing an exponent $n$ through
\begin{equation}
\label{n}
f(p) \sim (p - c)^n, \;\;\;\; p \to c,
\end{equation}
phase separation occurs iff $n > 0$. At the critical point 
$\rho = \rho_c$ the disorder averaged headway distribution has a power
law tail $\sim u^{-(n+2)}$ \cite{kf}. 
Evans \cite{evans96,evans97} has emphasized
the close analogy to Bose-Einstein condensation, where $f(p)$ plays
the role of a density of states, and the slowest particle in the system
corresponds to the quantum mechanical ground state. 

\subsection{Coarsening behavior}
\label{ParCoarse}

No exact results pertaining to the dynamics of phase separation are
available, apart from the observation \cite{timo99}
that the existence of a 
well-defined hydrodynamic limit implies that inhomogeneities 
are restricted to scales smaller than $t$, and therefore 
\begin{equation}
\label{sublinear}
\lim_{t \to \infty} \xi(t) / t = 0.
\end{equation}
Considerable evidence has however accumulated in favor of the idea
\cite{kf} that the coarsening behavior for particlewise disorder
can be described in terms of a simpler, deterministic model,
in which particles move ballistically on the real line with 
fixed random speeds and coalesce upon overtaking. Such a model was
first introduced by Newell \cite{newell}, and later a detailed kinetic 
theory was worked out by Ben-Naim, Krapivsky and Redner \cite{bennaim94}. 

Within the deterministic model, the dynamic exponent
$z$ can be determined through a simple extremal statistics argument.
The key idea is that the particles heading the platoons at time $t$
are those with the smallest speeds among of the order of $\xi(t)$
particles. Elementary probability theory suffices to show that,
for a probability density behaving as (\ref{n}), these extremal speeds
cluster in an interval of size $\xi^{-1/(n+1)}$ above the minimal speed
$c$. Therefore the speed difference $\Delta v$ between two platoons
is of the order $\xi^{-1/(n+1)}$, and the faster platoon will merge with
the slower one on a time scale $t \sim \xi/\Delta v \sim \xi^{(n+2)/
(n+1)}$. Inverting this relation one obtains the coarsening law 
(\ref{xi}) with
\begin{equation}
\label{z(n)}
z = \frac{n+2}{n+1}.
\end{equation}
Numerical results supporting (\ref{z(n)}) have been reported for
models with parallel update \cite{ktitarev97}, in simulations of 
jam dissolution \cite{timo99} and in a simulation study of 
a system with open boundaries \cite{bengrine99b}.   

\section{Sitewise disorder}
\label{Site}

\subsection{Disorder types}

We distinguish three cases which will turn out to represent different
classes of coarsening behavior. For {\em type I disorder} the dynamics
is totally asymmetric, $q(x) \equiv 0$, and the forward rates 
$p(x)$ are independent random variables in an interval $[c,1]$,
with a minimal rate $c > 0$. The simplest (and typical) example is that
of binary rates, with probability density
\begin{equation}
\label{binary}
f(p) = \phi \delta(p - c) + (1 - \phi) \delta(p - 1)
\end{equation}
where $\phi \in (0,1)$ denotes the fraction of slow sites. 
{\em Type II disorder} 
is similar to type I except that the support of the
probability density $f(p)$ extends all the way to $p=0$, i.e. the
minimal rate $c = 0$. For type II  disorder
nontrivial dynamics occurs only for
continuous $f(p)$. As in the models with particlewise disorder
(eq.(\ref{n})), the
important feature of $f(p)$ is the behavior near $p=0$, which can
be characterized by an exponent $n$ through the relation 
\begin{equation}
\label{n2}
f(p) \sim p^n, \;\;\; p \to 0.
\end{equation} 

Finally, for {\em type III} disorder not only the strength,
but also the direction of the bias is spatially random. A majority of
sites has a bias to the right, say, with $p(x) > q(x)$, while
a minority has $q(x) > p(x)$. If the one-dimensional lattice is
viewed as a transport path in a higher-dimensional disordered structure,
such as a percolation cluster, the 
stretches of minority sites  can be 
interpreted as ``backbends'' where the path turns back against the
direction of the driving field \cite{rama87}. Compared to the strong disorder
effects induced by the backbends, the randomness in the strength of the
bias is irrelevant. Therefore a representative example of type III disorder is
a model where the strength of the bias is constant, and only its
direction varies. This corresponds to setting $q(x) = 1 - p(x)$ and
choosing the $p(x)$ from a binary distribution which is symmetric
around $p = 1/2$, 
\begin{equation}
\label{binary2}
f(p) = (1 - \phi) \delta(p - b) + \phi \delta(p - (1-b)).
\end{equation}
Here $b\in (1/2,1)$ denotes the strength of the bias and 
$\phi \in (0,1/2)$ the fraction of minority sites.

It is easy to see that for type II and III disorder 
the stationary particle 
current vanishes in the infinite system limit, due to 
the existence of arbitrarily large stretches of arbitrarily small
jump rates (for type II) or arbitrarily long backbends (for type III).
As a consequence phase separation occurs
at any density $\rho \in (0,1)$, i.e. $\rho_c^- = 0$ and
$\rho_c^+ = 1$. For type I disorder the existence of a nontrivial
current function $J(\rho) > 0$ describing the large scale dynamics
of density profiles has been rigorously established, and it has
been shown that $J(\rho)$ is convex in the sense that $J''(\rho) \leq
0$ \cite{timo99b}. However, in contrast to the models with particlewise
disorder the stationary state is not known, and therefore an 
explicit computation of $J(\rho)$ is not possible. In the next section
some bounds on $J(\rho)$ will be derived and used to bound the
critical densities for type I disorder. The coarsening dynamics
for all three cases will be addressed in Section \ref{SiteCoarse}.
 
\subsection{Bounds on the critical density for type I disorder}

We first collect some obvious properties of $J(\rho)$. Due to
particle-hole symmetry we have $J(\rho) = J(1 - \rho)$. The current
is bounded from below by the current $c \rho(1 - \rho)$ of a pure
system with all rates equal to the minimal rate $c$, and from above
by the current $\rho(1 - \rho)$ of the system with all rates equal to
unity. A more precise upper bound is obtained by observing that
in the infinite system there are arbitrarily large stretches with
rates arbitrarily close to $c$. The maximum current that can be 
driven through such a stretch is $c/4$, the maximum value of 
$c \rho(1 - \rho)$. We conclude that
\begin{equation} 
\label{Jbound1}
c \rho(1 - \rho) \leq J(\rho) \leq 
\min[c/4, \rho(1 - \rho)].
\end{equation}

Numerical simulations of site-disordered exclusion models
\cite{tripathy98,csahok94,bengrine99} and related
growth models \cite{jk92,jk95} indicate that the upper
bound $c/4$ is attained in a finite density interval around $\rho = 
1/2$, which coincides with the phase separation interval 
$[\rho_c^-, \rho_c^+]$; by particle-hole symmetry $\rho_c^- = 
1 - \rho_c^+ \equiv \rho_c$. 
In the following our strategy will be to 
derive optimal lower and upper bounds $J_<(\rho)$ and $J_>(\rho)$
on the stationary current, which are then translated into 
lower and upper bounds
$\rho_c^<$, $\rho_c^>$ on $\rho_c$ through the relation
\begin{equation}
\label{rhobounds}
J_>(\rho_c^<) = J_<(\rho_c^>) = c/4.
\end{equation}
The lower current bound in (\ref{Jbound1}) does not give rise to
any nontrivial density bound, while the upper bound $\rho(1 - \rho)$
yields
\begin{equation}
\label{rholow1}
\rho_c \geq (1 - \sqrt{1 - c})/2.
\end{equation} 

For the case of binary disorder (eq.(\ref{binary}))
an improved lower bound on the current was derived by Tripathy and
Barma \cite{tripathy98} by considering a finite ring of $L$ sites, 
$N = \rho L$ particles and $N_s = \phi L$ slow sites. They start from
the observation that the maximum current that can be driven through
a stretch of slow sites is a decreasing function of the length 
of the stretch (we will return to this point below in Section
\ref{SiteCoarse}). It is therefore plausible (though not rigorously
established) that for given $L$, $N$ and $N_s$ the stationary
current will be minimal in the {\rm fully segregated} limit where
all slow sites form a single large stretch. For $L \to \infty$
the fully segregated system can be treated as two 
connected homogeneous systems with different densities, which are
fixed through the constraints of equal currents and total particle
number. This yields the upper density bound
\begin{equation}
\label{FSS}
\rho_c \leq (1 - (1 - \phi)\sqrt{1 - c})/2.
\end{equation} 
In the dilute limit $\phi \to 0$ the bounds (\ref{rholow1}) and (\ref{FSS}) coincide,
and give $\rho_c = (1 - \sqrt{1 - c})/2$ exactly. It should however be noted
that this limit does {\em not} correspond to the case of a single defect site,
since the maximal current that can be driven through a single defect is larger 
than $c/4$ \cite{jano92} (see also Section \ref{SiteCoarse}). 

The lower bound (\ref{rholow1}) can be improved by comparing the
disordered exclusion model to a {\em zero range process} (ZRP)
with the same set of jump rates $\{p(x)\}$. In the ZRP
an arbitrary number of
particles is allowed on any site \cite{spitzer,krug97}, and therefore
any attempted jump succeeds. As a consequence the stationary state
of the ZRP is a product measure, with 
the occupation numbers at different sites
being independent, for any choice of jump rates 
depending on the position 
$x$ and on the number of particles at the site \cite{benjamini96,andjel82}.
Here we consider the case where the rate at which a particle is
transferred from site $x$ to $x+1$ is equal to $p(x)$ independent
of the number of particles at $x$, provided the latter is not zero. 
It is then obvious (and can be proved through waiting time considerations)
that the particle current $J_{\rm ZRP}(\rho)$ 
of the ZRP provides an upper bound
to the current $J(\rho)$ of the ASEP. 

In fact the disordered ZRP is equivalent to the ASEP with 
{\em particlewise} disorder, with the ZRP occupation numbers 
representing the headways in the ASEP 
\cite{benjamini96,krug97}. The ZRP current is equal to the
particle speed $v$ of the ASEP, which is given by (\ref{v}) for
any disorder distribution $f(p)$. The ZRP density is equal to 
the mean headway of the ASEP, and is therefore related to the ASEP
density through $\rho_{\rm ZRP} = 1/\rho_{\rm ASEP} - 1$.
Evaluating the integral in (\ref{v}) for the binary distribution
(\ref{binary}) yields
\begin{equation}
\label{JZRP}
\rho_{\rm ZRP} = J_{\rm ZRP}
\left(\frac{\phi}{c - J_{\rm ZRP}} + \frac{1 - \phi}{1 - J_{\rm ZRP}}
\right),
\end{equation} 
and setting $J_> = J_{\rm ZRP}$ in (\ref{rhobounds}) we obtain the
density bound
\begin{equation}
\label{rhozrp}
\rho_c \geq \frac{\phi}{3} + \frac{c(1 - \phi)}{4 - c},
\end{equation}
which improves (\ref{rholow1}) for small $c$. In particular,
for $c \to 0$ we have $\rho_c \geq \phi/3$, which proves,
remarkably, that
the homogeneous phase $\rho < \rho_c$ persists even when the
slow sites become complete blockages. In Figure 3 the bounds
(\ref{rholow1}), (\ref{FSS}) and (\ref{rhozrp}) are compared
to numerical data.

\subsection{Coarsening behavior}
\label{SiteCoarse}

At least for type I disorder the existence of a hydrodynamic
limit \cite{timo99b}
implies that the relation (\ref{sublinear}) carries over
to the sitewise case. To obtain a finer estimate of the coarsening
scale $\xi(t)$ we rely on extremal statistics arguments similar to
those used in Section \ref{ParCoarse}. A schematic 
phase separated density profile is shown in Figure 4. 
Two ``antischocks'' at positions $x_1$ and 
$x_2$, where the density jumps from 
$\rho_c^+ = 1 - \rho_c$ to $\rho_c^- = \rho_c$, mark 
{\em bottleneck} regions of particularly slow rates,
which support maximum currents $j_1$ and $j_2$. If the bottleneck
in the downstream direction is slightly more restrictive, in the sense
that $\Delta j = j_1 - j_2 > 0$, 
then the low density region between the bottlenecks
will slowly fill in and disappear at a time 
\begin{equation}
\label{bottleneck}
t \approx (1 - 2 \rho_c) \xi/\Delta j.
\end{equation}
If the statistics of extremal bottlenecks is known, the
typical current difference $\Delta j$ can be estimated as a function
of $\xi$ and (\ref{bottleneck}) yields a prediction for the coarsening
law $\xi(t)$. In the following this will be carried out for the 
different disorder types.

\subsubsection{Type I disorder}

Consider first the conceptually simplest case of the binary
disorder distribution (\ref{binary}). We expect long stretches
of slow sites to constitute the most restrictive bottlenecks. 
For a quantitative analysis we would require the maximum current
$j_{\rm max}(c, \ell)$ which can be driven through a stretch
of $\ell$ slow sites with jump rates $c$ embedded in an infinite
system of sites with jump rates 1. Already for $\ell = 1$
the computation of $j_{\rm max}(c, \ell)$ is a difficult unsolved
problem \cite{jano92}. However for large $\ell$ we can make
progress by replacing the stretch 
by a finite system of $\ell$ sites with uniform jump rates $c$
and periodic or open boundary conditions, for which the maximum
current is known \cite{schuetz93,hakim93}. For both kinds
of boundary conditions the current approaches the $\ell \to \infty$
limit $c/4$ from above, with a leading correction 
\cite{jkmeakin} proportional to
$1/\ell$. Thus we expect, for large $\ell$, 
\begin{equation}
\label{jmax}
j_{\rm max}(c, \ell) \approx (c/4)(1 + a/\ell) + 
{\cal O}(1/\ell^2),
\end{equation}
where $a$ is a positive constant of order unity.  

Since the probability distribution of the lengths of slow stretches 
is 
\begin{equation}
\label{Pl}
P(\ell) = (1 - \phi)\phi^\ell,
\end{equation} 
the longest stretch in a region
of size $\xi$ is of the order of 
\begin{equation}
\label{lmax}
\ell_{\rm max} \approx \frac{\ln \xi}{\ln (1/\phi)}.
\end{equation}
Note that $\ell_{\rm max} \ll \xi$, which is consistent with the
assumption of well-localized bottlenecks inherent in Figure 4.
Using (\ref{jmax}) we see that the currents supported by the
longest stretches exceed $c/4$ by an amount of the order of
$c/\ell_{\rm max}$, and therefore 
\begin{equation}
\label{deltaj}
\Delta j \approx \frac{c \ln (1/\phi)}{\ln \xi}.
\end{equation}
Inserting this into (\ref{bottleneck}) the 
leading order coarsening law is obtained as
\begin{equation}
\label{xiI}
\xi(t) \sim \frac{t/t_0}{\ln (t/t_0)}
\end{equation}
with a characteristic time scale $t_0 \sim (1 - 2 \rho_c)/c \ln(1/\phi)$.
This argument was formulated earlier in the context of phase-disordered
growth models, where also numerical evidence in favor of the coarsening
law (\ref{xiI}) was presented \cite{jk92,jk95}.  

For continuous disorder distributions $f(p)$ the identification of 
the relevant bottlenecks is a little more subtle. Consider
a region of size $\ell$ where all rates satisfy $p(x) \leq c + \epsilon$.
The maximum current through such a region can then be estimated as
\begin{equation}
\label{jmaxcont}
j_{\rm max} \approx \frac{c + \epsilon}{4}\left(1 + \frac{a}{\ell} \right)
\approx c/4 + \epsilon/4 + ca/4 \ell \equiv c/4 + j.
\end{equation} 
If $f(p)$ behaves as in (\ref{n}) for $p \to c$ the probability of the
region is of the order of $\epsilon^{(n+1)\ell}$. The probability distribution
of $j$ can then be written as
$$
P(j) \sim \int d\ell \int d\epsilon \; \epsilon^{(n+1)\ell}
\delta(j - (\epsilon + ca/\ell)/4) \sim
$$ 
\begin{equation}
\label{jmaxprob}
\int d\epsilon \exp[-(n+1) ca \ln(1/\epsilon)/(4 j - \epsilon)].
\end{equation}
Evaluating the last integral at the saddle point yields 
\begin{equation}
\label{P(j)}
P(j) \sim \exp[-(n+1) ca \ln^2(1/j)/j].
\end{equation}
The current scale $\Delta j$ of the most restrictive among $\xi$ bottlenecks
is obtained by setting $P(\Delta j) \sim 1/\xi$, which gives 
\begin{equation}
\label{deltaj2}
\Delta j \sim \frac{[\ln(\ln \xi)]^2}{\ln \xi},
\end{equation}
and the corresponding coarsening law reads, to leading
order in $t$,
\begin{equation}
\label{xiII}
\xi(t) \sim \frac{t [\ln(\ln t)]^2}{\ln t}
\end{equation}
which, for most purposes, is indistinguishable from
(\ref{xiI}).

\subsubsection{Type II and III disorder}

For type II disorder with a continuous probability distribution
$f(p)$, characterized by 
(\ref{n2}), 
the expression (\ref{jmaxcont}) for the maximum current
supported by a slow stretch of length $\ell$ 
applies with  $c=0$. The distribution of $j_{\rm max}$  then 
becomes 
$$
P(j_{\rm max}) \sim \int d\ell \int d\epsilon \; \epsilon^{(n+1)\ell}
\delta(j_{\rm max} - \epsilon(1 + a/\ell)/4) \sim
$$ 
\begin{equation}
\label{jmaxprob2}
\int d\ell \exp[-(n+1) \ell \ln((1 + a/\ell)/4 j_{\rm max})].
\end{equation}
Now the maximum of the exponent evidently occurs at $\ell = 1$, i.e.
the dominant bottlenecks are {\em individual slow sites}. The distribution
of the currents supported by the bottlenecks is then simply given by
the jump rate distribution $f(p)$ itself, and the situation reduces
to that analyzed in the case of particlewise disorder, Section 
\ref{ParCoarse}. In particular, the coarsening exponent $z$ for 
type II sitewise disorder is also given by (\ref{z(n)}).

For type III disorder with distribution
(\ref{binary2}) the dominant bottlenecks are long backbends,
i.e. stretches of minority sites at which the local bias
is directed against the mean flow direction. 
The maximum current that can be
driven through a backbend of length $\ell$ is exponentially small 
in $\ell$, and is given by \cite{tripathy98,rama87}
\begin{equation}
\label{jbackbend}
j_{\rm max}(\ell) \sim \exp[-(1/2) \ell \ln(b/(1-b))].
\end{equation}
Combining this with the probability distribution (\ref{Pl})
of backbend lengths it follows that $j_{\max}$ is distributed
according to a power law,
\begin{equation}
\label{jmaxIII}
P(j_{\rm max}) \sim (j_{\rm max})^{2 \theta^{-1} - 1}
\end{equation}
where 
\begin{equation}
\label{theta}
\theta = \frac{\ln[b/(1-b)]}{\ln[1/\phi]}.
\end{equation}
Since the largest backbend in a region of size $\xi$ is of length
$\ell \sim \ln \xi \ll \xi$, we can employ a coarse grained 
picture in which the backbends are shrunk to individual sites
with a jump rate distribution given by (\ref{jmaxIII}), thus 
effectively reducing the problem to type II disorder with the
exponent $n$ in (\ref{n2}) given by $n = 2/\theta - 1$. The coarsening
exponent for the disorder distribution (\ref{binary2}) is then
obtained from (\ref{z(n)}) as
\begin{equation}
\label{zIII}
z = 1 + \theta/2.
\end{equation}

\subsection{Relation to directed polymers}
\label{DP}

Using the waiting time approach \cite{tang94}
the site disordered ASEP can be mapped
to a zero temperature directed polymer (DP) with point and columnar
disorder \cite{thh94}. In that context the coarsening law 
$\xi(t)$ describes the disorder-induced transverse wandering
of the polymer, which can be estimated using variable range hopping
arguments \cite{thh94} and the analogy to Lifshitz tails for one-dimensional
disordered Schr\"odinger operators \cite{thh93}. 

To see that the
results derived for the DP are consistent with those obtained above,
it is important the recall \cite{tang94} that the waiting time
mapping transforms the time $t$ of the ASEP into the {\em energy}
of DP. For type I disorder the transverse wandering $\delta x$
of the DP was
found to increase with its length $L$ as \cite{thh94}
\begin{equation}
\label{dx1}
\delta x \sim L/(\ln L)^2,
\end{equation}
while the ground state energy behaves as $E \sim L/\ln L$ to leading
order. Combining the two results and identifying $E \sim t$ the
coarsening law (\ref{xiI}) follows. 

For type II disorder the power law (\ref{n2}) of the probability
distribution at small $p$ translates into a power law tail
\begin{equation}
\label{Ptau}
P(\tau) \sim \tau^{-(2 + n)}, \;\;\; \tau \to \infty
\end{equation}
in the distribution of waiting times or energies $\tau = 1/p$.
Directed polymers in the presence of columnar disorder with
a power law distribution were considered in Ref.\cite{thh93},
where it was shown that the wandering is typically ballistic,
$\delta x \sim L$, while the ground state energy scales with
length as 
\begin{equation}
\label{EL}
E \sim L^{(n+2)/(n+1)}
\end{equation}
in agreement with (\ref{z(n)}).
It is worthwhile to point out that in the DP context the scaling
laws (\ref{dx1}) and (\ref{EL}) were also confirmed numerically
\cite{thh94,thh93}. 

\section{Summary and open questions}
\label{Conclusions}

In this paper I have described some recent progress in our
understanding of disorder effects in asymmetric simple exclusion
models. A common feature of both
particlewise and sitewise disordered systems is the appearance
of phase separation in an interval of densities, which is
{\em macroscopically} characterized by a linear portion in 
the current-density relation $J(\rho)$; in the particlewise case
$J(\rho) = c \rho$ for $\rho < \rho_c$, while in the sitewise
case $J(\rho) \equiv c/4$ for $\rho_c \leq \rho \leq 1 - \rho_c$.
An interesting open question concerns the connection between
phase separation and linearity of $J(\rho)$, which is reminiscent
of the role that the convexity of thermodynamical potentials plays
for the stability of equilibrium systems. While it is obvious
that phase separation implies a linear segment in $J(\rho)$, the converse
statement has, to my knowledge, not been established. To prove
that it is false, it would be sufficient to find a (noisy!) exclusion
type model with a homogeneous stationary state and a linear current-density
relation (deterministic systems with linear $J(\rho)$ are well known
\cite{nagel96,nagel93}). 

The dynamics of phase separation has been explored in the framework
of scaling arguments, which can be formulated in a similar way both
for particlewise and sitewise disorder. In the particlewise case
the relevant bottlenecks which determine the positions of domain
boundaries are always individual slow particles, while in the
sitewise case with type I and III disorder the bottlenecks are formed
collectively by many defects. For type I disorder this implies a 
certain {\em universality} of the coarsening law, in the sense
that the exponent $z$ in (\ref{xi}) is $z=1$ independent of the
underlying disorder distribution; the additional logarithmic
corrections in (\ref{xiI},\ref{xiII}) ensure the consistency
with the rigorous result (\ref{sublinear}). This is somewhat
analogous to the case of {\em finite temperature}
directed polymers with columnar defects, where universal
scaling laws arise from the thermal averaging over large spatial
regions \cite{thh93}.

A numerical confirmation of the predictions for the coarsening
dynamics in the case of sitewise disorder would be most welcome.
For type II disorder this should be relatively straightforward,
however in the cases of type I and III disorder the behavior is 
dominated by exponentially rare regions, which may make it hard to
reach asymptopia.

\vspace{0.5cm}

{\bf Acknowledgements.} The results on particlewise disorder were
obtained in collaboration wit Pablo Ferrari and Timo 
Sepp\"al\"ainen, while the material on sitewise disorder is joint
work with Mustansir Barma and Goutam Tripathy. 
Support by DAAD within the PROBRAL programme
and by DFG within SFB 237 is gratefully acknowledged.

\newpage

\pagestyle{empty}

\begin{figure}
\centerline{\leavevmode
\epsfxsize=12cm
\epsfbox{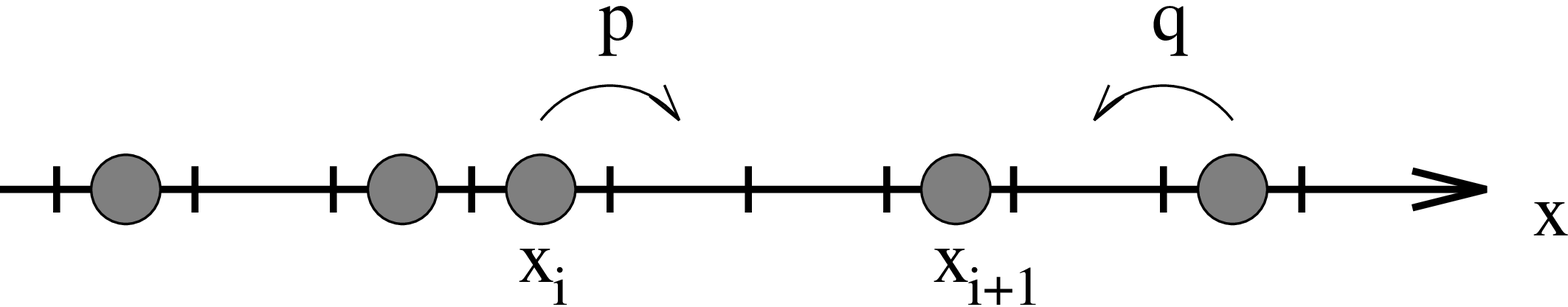}}
\caption{Illustration of the asymmetric simple exclusion process.}
\label{asep}
\end{figure}

\begin{figure}[htb]
\setlength{\unitlength}{0.00333333333333truein}
\begin{center}
\begin{picture}(1255,2400)
\Large
\thicklines
\put(135,2530){\vector(1,0){1070}}
\put(647,2530){\line(0,1){20}}
\put(1160,2530){\line(0,1){30}}
\put(1030,2580){$1024$}
\put(1220,2520){$x$}
\put(135,2530){\vector(0,-1){2450}}
\put(135,2030){\line(-1,0){20}}
\put(135,1530){\line(-1,0){30}}
\put(-20,1515){1000}
\put(135,1030){\line(-1,0){20}}
\put(135,530){\line(-1,0){30}}
\put(-20,515){2000}
\put(100,0){$t$}
\put(80,2580){0}
\put(150,135){\epsfbox{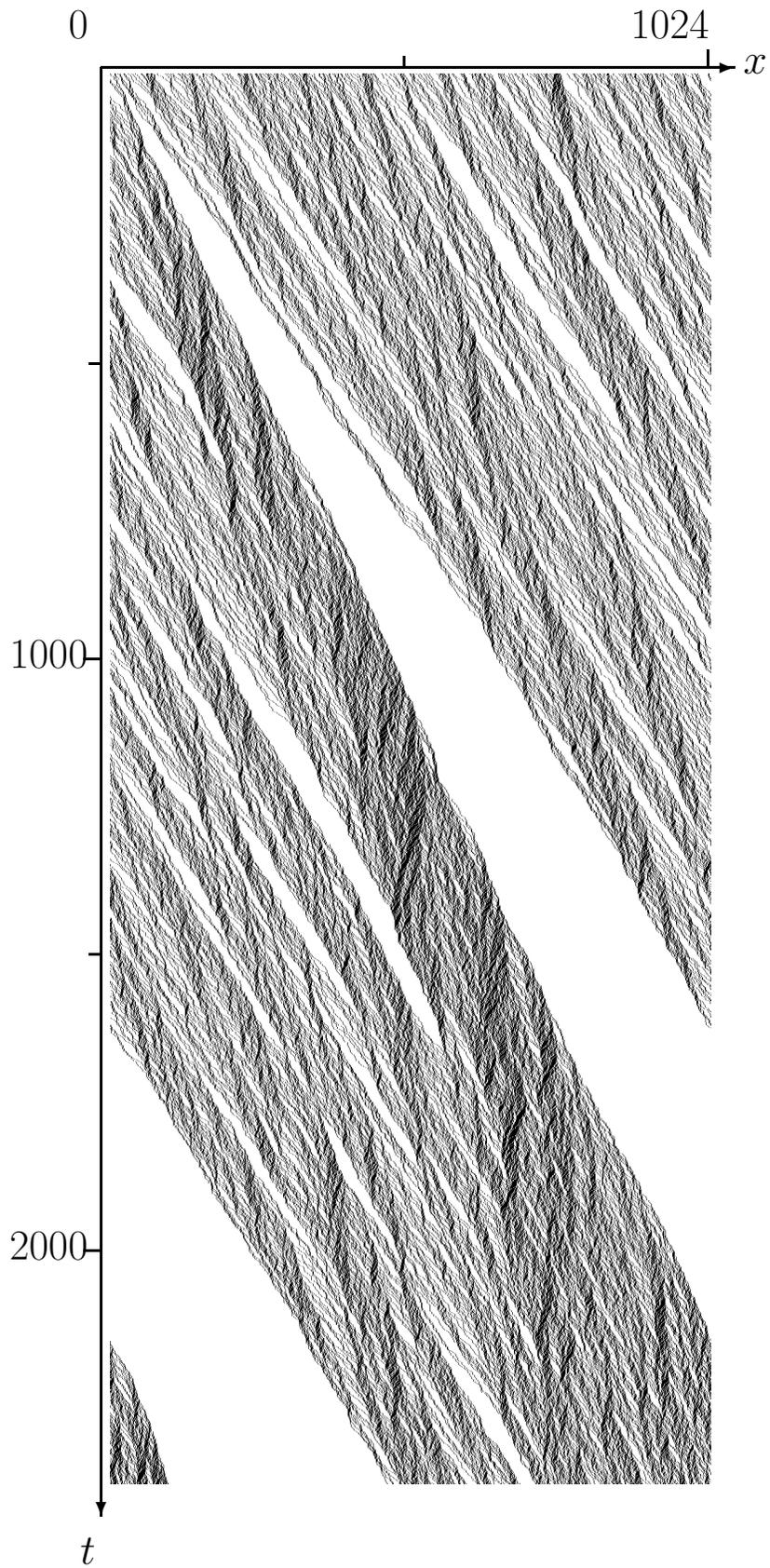}}
\end{picture}
\end{center}
\normalsize
\caption[f2] {Space-time plot of trajectories of the ASEP with particlewise
disorder. The figure shows 256 particles on a ring of 1024 sites. The
initial distribution of particles was random with density $\rho = 1/4 < 
\rho_c = 0.4$. Courtesy of M. Gerwinski.}
\end{figure}

\begin{figure}
\centerline{\leavevmode
\epsfxsize=10cm
\epsfbox{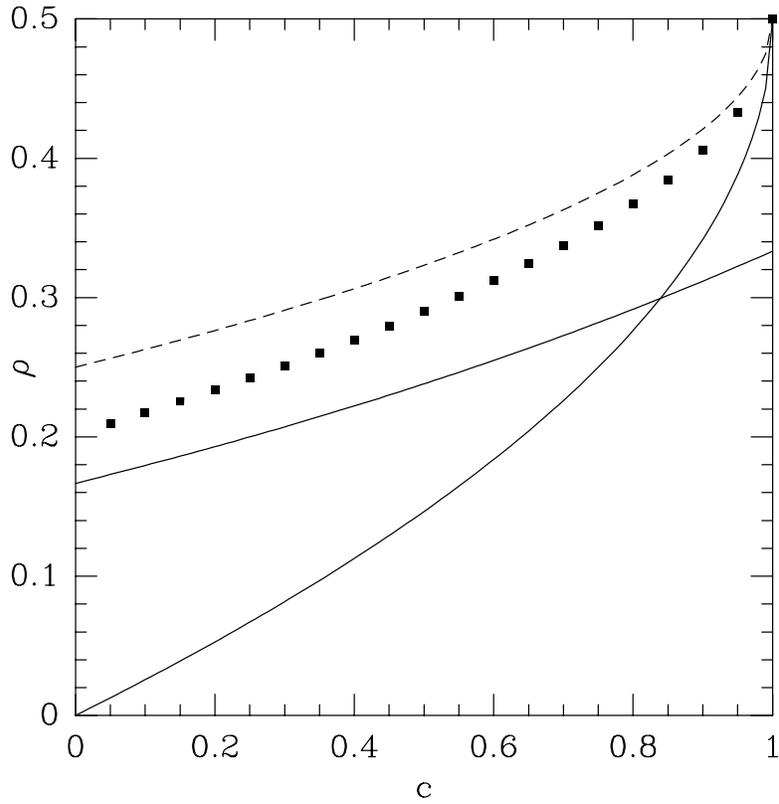}}
\caption{Bounds on the critical density for sitewise disorder with
the binary distribution (\ref{binary}), $\phi = 1/2$. The dashed line
is the upper bound (\ref{FSS}), the two full curves show the
lower bounds (\ref{rholow1}) and (\ref{rhozrp}), and the full squares
are simulation data obtained by G. Tripathy.}
\label{bounds}
\end{figure}

\begin{figure}
\centerline{\leavevmode
\epsfxsize=12cm
\epsfbox{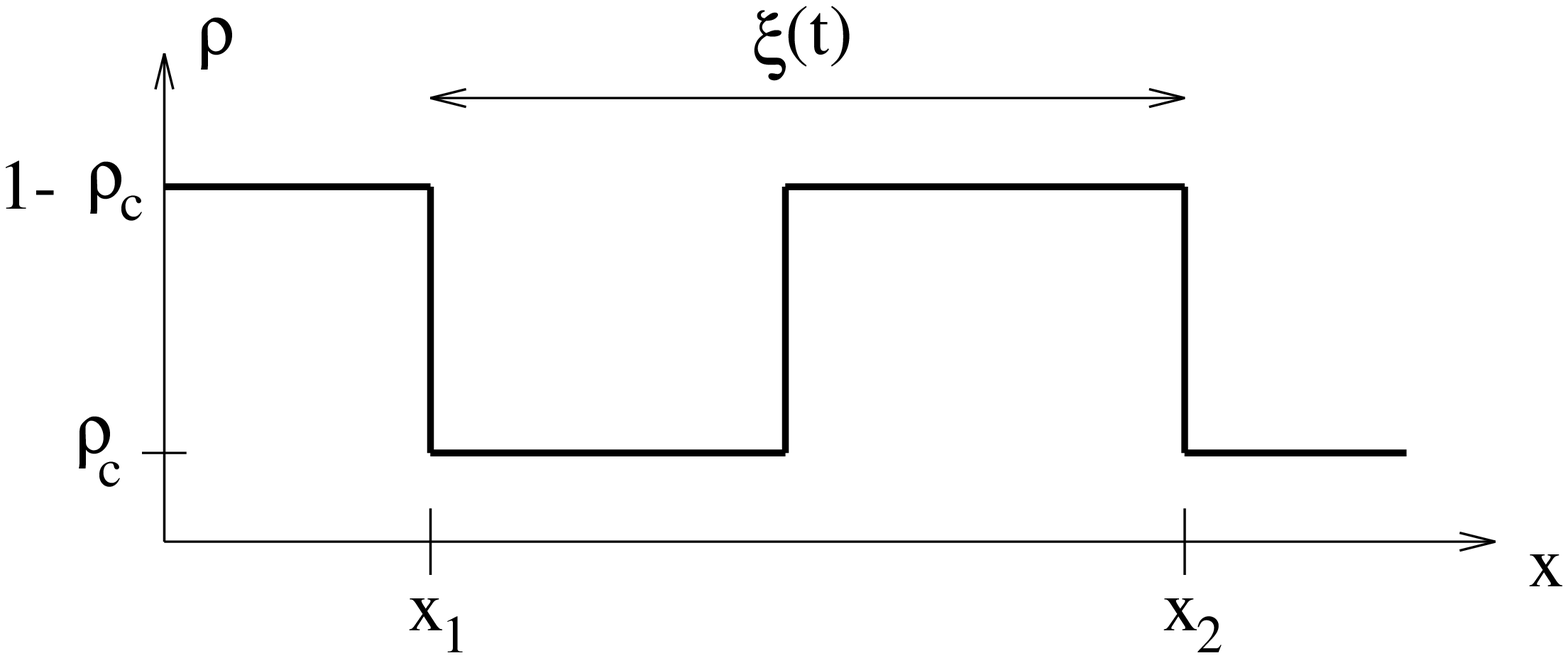}}
\caption{Schematic of two bottlenecks at positions $x_1$ and $x_2$.}
\label{bottle}
\end{figure}


\begin{thebibliography}{99}

\bibitem{spitzer} F. Spitzer, 
Adv. in Math. {\bf 5}, 256 (1970). 
\bibitem{liggett} T.M. Liggett, {\em Interacting Particle Systems}
(Springer, Berlin 1985). 
\bibitem{liggett99} T. Liggett, {\em Stochastic Interacting Systems: Contact,
Voter and Exclusion Processes} (Springer, Berlin/Heidelberg 1999). 
\bibitem{spohn} H. Spohn, {\em Large Scale Dynamics of Interacting Particles}
(Springer, Berlin/Heidelberg 1991).
\bibitem{kipnis99} C. Kipnis and C. Landim, {\em Scaling Limits of Interacting
Particle Systems} (Springer, Berlin/Heidelberg 1999). 
\bibitem{schuetz97} G. Sch\"utz, Int. J. Mod. Phys. B {\bf 11}, 197
(1997), and references therein.
\bibitem{nagel96} K. Nagel,\PRE{53}, 4655 (1996).
\bibitem{krug91}J. Krug, \PRL{67}, 1882 (1991).
\bibitem{schuetz93} G. Sch\"utz and E. Domany, \JSP{72}, 277 (1993).
\bibitem{hakim93} B. Derrida, M. R. Evans, V. Hakim, and V. Pasquier, 
\JPA{26}, 1493 (1993).
\bibitem{ks91} J. Krug and H. Spohn, in 
{\em Solids Far From Equilibrium}, ed. by C. Godr\`eche
(Cambridge University Press, Cambridge 1991), p. 479.
\bibitem{hhz95} T. Halpin-Healy and Y.C. Zhang, Phys. Rep. {\bf 254}, 215 
(1995).
\bibitem{newell} G.F. Newell, Opns. Res. {\bf 7}, 589 (1959).
\bibitem{bennaim94} E. Ben-Naim, P. Krapivsky, S. Redner,\PRE{50}, 822 (1994).
\bibitem{krug97} J. Krug, in ``Traffic and Granular Flow '97' '', 
ed. by M. Schreckenberg and D. E. Wolf (Spinger, Singapore 1998), p. 285.
\bibitem{csahok94} Z. Csahok and T. Vicsek, \JPA{27}, L591 (1994).
\bibitem{knospe97} W. Knospe, L. Santen, A. Schadschneider and M. Schreckenberg,
in ``Traffic and Granular Flow '97' '', 
ed. by M. Schreckenberg and D. E. Wolf (Spinger, Singapore 1998), p. 349.
\bibitem{harms97} T. Harms and R. Lipowsky, \PRL{79}, 2895 (1997).
\bibitem{benjamini96} I. Benjamini, P.A. Ferrari and C. Landim,
Stoch. Proc. Appl. {\bf 61}, 181 (1996).
\bibitem{kf}
J. Krug and P.A. Ferrari,
J. Phys. A {\bf 29}, L465 (1996). 
\bibitem{evans96} M.R. Evans, 
Europhys. Lett. {\bf 36}, 13 (1996).
\bibitem{evans97} M.R. Evans,
J. Phys. A {\bf 30}, 5669 (1997). 
\bibitem{timo99} T. Sepp\"al\"ainen and J. Krug, J. Stat. Phys. {\bf 95},
525 (1999).
\bibitem{tripathy98} G. Tripathy and M. Barma,  
\PRE{58}, 1911 (1998).
\bibitem{bengrine99} M. Bengrine, A. Benyoussef, H. Ez-Zahraouy and 
F. Mhirech, Phys. Lett. A {\bf 253}, 135 (1999).
\bibitem{ktitarev97} D. V. Ktitarev, D. Chowdhury and D. E. Wolf, \JPA{30}, L221 (1997).
\bibitem{bengrine99b} M. Bengrine, A. Benyoussef, H. Ez-Zahraouy,
J. Krug, M. Loulidi and F. Mhirech, \JPA{32}, 2527 (1999).
\bibitem{rama87} R. Ramaswamy and M. Barma, \JPA{20}, 2973 (1987).
\bibitem{timo99b} T. Sepp\"al\"ainen, Ann. Probab. {\bf 27}, 361 (1999).
\bibitem{jk92} J. Krug, in ``Surface Disordering: Growth, Roughening and
Phase Transitions'', ed. by R. Jullien, J. Kert\'esz, P. Meakin and D.E. Wolf
(Nova Science, 1992), p. 177.
\bibitem{jk95} J. Krug, \PRL{75}, 1795 (1995).
\bibitem{jano92} S. A. Janowsky and J. L. Lebowitz, \PRA{45}, 618 (1992);
S.A. Janowsky and J.L. Lebowitz, \JSP{77}, 35 (1994).
\bibitem{andjel82} E.D. Andjel, Ann. Probab. {\bf 10}, 525 (1982).
\bibitem{jkmeakin} J. Krug and P. Meakin, \JPA{23}, L987 (1990).
\bibitem{tang94} J. Krug and L.-H. Tang, \PRE{50}, 104 (1994).
\bibitem{thh94} I. Arsenin, T. Halpin-Healy and J. Krug,
\PRE{49}, R3561 (1994).
\bibitem{thh93} J. Krug and T. Halpin-Healy, J. Physique I France 
{\bf 3}, 2179 (1993).
\bibitem{nagel93} K. Nagel and H.J. Herrmann,
Physica A {\bf 199}, 254 (1993).


\end{thebibliography}
\end{document}